\documentclass[%
 reprint,
 amsmath,amssymb,
 aps,
showpacs
]{revtex4-2}

\usepackage{graphicx}
\usepackage{dcolumn}
\usepackage{bm}

\usepackage{xcolor}

\usepackage{graphics, amsmath,amssymb,multirow}
\begin{document}
\pacs{47.57.E-, 87.19.U-, 47.55.Kf, 87.85.gf}
\preprint{APS/123-QED}
\title{Axial dispersion of red blood cells in microchannels}

\author{Sylvain Losserand}
\email{losserand.sylvain@gmail.com}
\author{Gwennou Coupier}%
\email{gwennou.coupier@univ-grenoble-alpes.fr}
\affiliation{Universit\'e Grenoble Alpes, CNRS, LIPhy, F-38000 Grenoble, France}%


\author{Thomas Podgorski}
\email{thomas.podgorski@univ-grenoble-alpes.fr}
\affiliation{
Universit\'e Grenoble Alpes, CNRS, Grenoble INP, LRP, F-38000 Grenoble, France
}
\altaffiliation[Formerly at]{
Universit\'e Grenoble Alpes, CNRS, LIPhy, F-38000 Grenoble, France
}

\date{\today}

%
%
%
%
\begin{abstract}
Red blood cells flowing in a microchannel undergo dispersion in the flow direction due to the non-uniform velocity profile while transverse migration due to  {flow-induced deformations of cells combined with the presence of walls and a parabolic velocity profile} 
tends to focus them along the center line. This results in a dispersion of RBC transit times through a capillary that is directly related to their transverse migration properties. By analogy with the Taylor-Aris problem, we present an experimental method to characterise this phenomenon by injecting pulses of  {dilute suspensions of} red blood cells and measuring the evolution of their length along the channel, and varying mechanical parameters such as RBC deformability and fluid viscosity. A direct comparison of experimental results with a model that incorporates longitudinal advection and transverse migration  {in the dilute limit} shows that this principle provides through a simple dispersion measurement an evaluation of migration characteristics that are directly connected to cell mechanical properties.

\end{abstract}

%
%

\maketitle

\section{ Introduction} \label{sec:intro}

Blood is a typical example of a suspension of deformable particles whose flow is intimately governed by the mechanics of Red Blood Cells (RBCs), which make up about 45\% of blood volume. Their deformability is responsible for specific dynamic behaviors in flow that induce significant differences in hydrodynamic properties compared to generic suspensions of rigid particles and control the structure and rheology of the suspension in confined channel flow. For instance, RBCs, vesicles {, drops} or elastic capsules experience migration forces that usually drive them away from walls towards the center of the channel  {\cite{Abkarian02,popel05,callens08,coupier08,geislinger12,hariprasad12,losserand19,smart91,griggs07}. This transverse migration, that exists even without inertial effects, is a consequence of the shear induced deformations of particles that breaks the fore-aft symmetry. Both the presence of walls \cite{Abkarian02,callens08,smart91,grandchamp13} and curvature of the velocity profile \cite{coupier08,geislinger12,griggs07} contribute to the migration of deformable particles in channel flow. Similarly,} hydrodynamic interactions and pair-collisions between  {deformable} particles lead to transverse shear-induced diffusion \cite{grandchamp13,gires14,Malipeddi2021,Loewenberg97,singh09,Malipeddi2019}. 
Both phenomena lead to transverse motions of particles in the non-uniform, Poiseuille-like velocity field of the channel flow which, at steady state leads to an inhomogeneous distribution of particles or cells, with consequences on the effective rheology \cite{Lindqvist31,fedosov14b,feng21,audemar2022} but also on mixing and dispersion \cite{Bishop02}, both in the transverse and axial directions.

Indeed, a well-known phenomenon in thin channel or tube flow of solutions or suspensions is the axial dispersion of molecules or particles due to the non-uniform velocity distribution. In the case of molecular solutions or suspensions of Brownian particles for which thermal diffusion is the major mechanism of exploration of streamlines by particles, this phenomenon is known as Taylor-Aris dispersion \cite{taylor53,aris56}. In this context, an effective diffusion coefficient in the axial direction can be derived by expressing the advection-diffusion equation in terms of small deviations from cross-sectionnaly averaged quantities.  {It depends on the Peclet number $Pe=a U/D$ that compares advection and diffusion effects, where $a$ is the tube radius, $U$ the cross-section averaged flow velocity and $D$ the diffusion constant}. Using a similar approach, Griffiths \& Stone derived the axial diffusion properties for colloidal suspensions of particles that experience shear-induced diffusion in addition to Brownian diffusion \cite{griffiths12} and showed that the process becomes essentially nonlinear and slower than pure Brownian-based Taylor-Aris dispersion. 

A direct consequence of the axial dispersion of particles is a dispersion of their transit times through a channel or channel network: particles entering the channel at the same time, e.g. by injecting a pulse at the inlet, may exit at very different times due to the spreading of the pulse. This is a well known limitation to the resolution of chromatographic techniques \cite{tabeling2005}. In blood microcirculation, the axial dispersion of RBCs leads to a dispersion of transit times through an organ, which may have an influence on oxygen release. This phenomenon was studied in a pioneering study in-vivo by Lipowsky \cite{lipowsky93} who showed that both the average transit time of a pulse of fluorescently labeled RBCs and the dispersion of this pulse was strongly influenced by the mechanical properties of RBCs. More specifically, artificially rigidified cells had significantly longer and more dispersed transit times. While part of the dispersion in capillary networks is inherent to the network's complex structure in which multiple paths are possible between the inlet and outlet, its dependence on cell rigidity suggests that axial dispersion in each individual capillary influences the overall behavior.

\begin{figure*}[t]

\resizebox{0.8\textwidth}{!}{
  \includegraphics{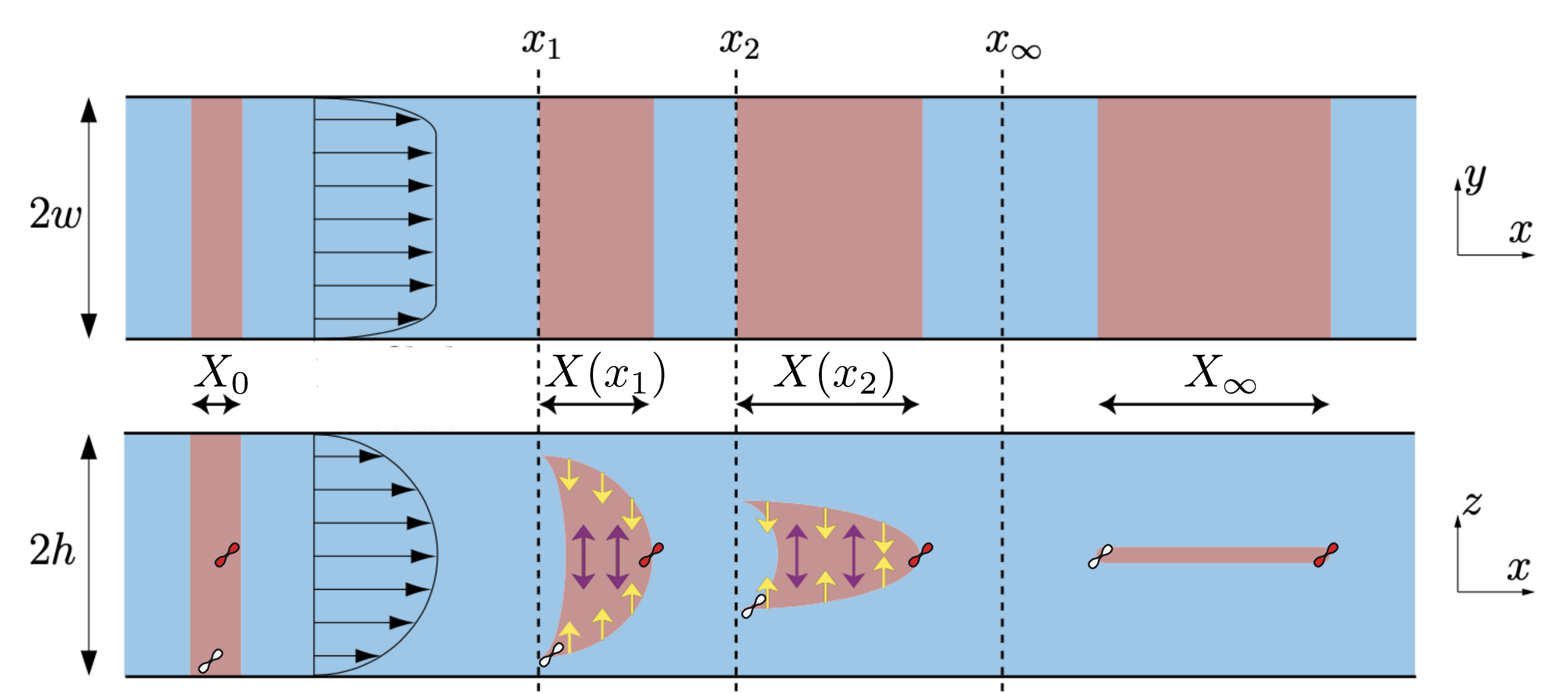}}\caption{Schematic view of the mechanisms leading to longitudinal diffusion for a pulse of non-Brownian particles in interactions with wall and with neighbors. The top panel shows the observed elongation of the pulse at different positions $x_i$ in the no-shear plane $xy$ while the bottom panel illustrates the underlying mechanism in the plane of shear. Cells that are initially centered are the fastest ones at the front of the pulse (moving at velocity $u_0$) while those starting close to walls are the slowest ones 
  and will constitute the tail of the distribution. As they are displaced in the $z$ direction due to transverse migration (yellow arrows) and possibly cell-cell interactions (purple arrows), a complex evolution of the time-lag between the front and the back of the pulse takes place along the channel until a stationary distribution of length $X_\infty$ is reached in the asymptotic regime. Note that in the case of polydisperse suspensions, segregation (both transverse and axial) may take place due to these dynamics. 
  \label{fig:expl}}
\end{figure*}

In contrast with the Taylor-Aris dispersion or the axial dispersion of colloidal suspensions in which the sole transverse motion mechanism is diffusive, whether it is Brownian or nonlinear, the transport of deformable particles in tube or channel flow is strongly influenced by migration towards the centerline, which tends to decrease the cross-sectional dispersion of cells over time {, especially in dilute suspensions}.  This convective effect decreases the axial dispersion rate as the centering of cells takes place (see Fig. \ref{fig:expl}). As lateral migration increases with parameters such as deformability or size of the particles \cite{coupier08,geislinger12,losserand19}, it is expected that cells having a faster lateral migration velocity in tube flow will undergo less axial dispersion, in a manner similar to changes in transit times in more complex networks \cite{lipowsky93}.

In this work, we made an experimental study of the axial dispersion of pulses of RBCs in a straight flat channel (2D Poiseuille flow) by varying mechanical parameters such as the deformability of these cells through population selection by density gradient separation and the viscosity of the suspending medium.  {For dilute suspensions,} the dispersion rates are related to previously established lateral migration laws of individual cells and an asymptotic theoretical model is proposed in which the evolution of the pulse length is explicitly related to RBC migration parameters and channel thickness.  {From this model we show that in the dilute limit a}  simple, macroscopic measurement of the axial dispersion of a pulse can be used to derive microscopic migration coefficients and is a marker of RBC deformability in a blood sample. Finally, we present results on the combined effect of axial dispersion in straight channels and the role of diverging and converging bifurcations in a simple channel network that provides insight on the mechanisms leading to the dispersion of transit times in complex microvascular networks.

\begin{figure}[t]

\resizebox{8.6cm}{!}{
  \includegraphics{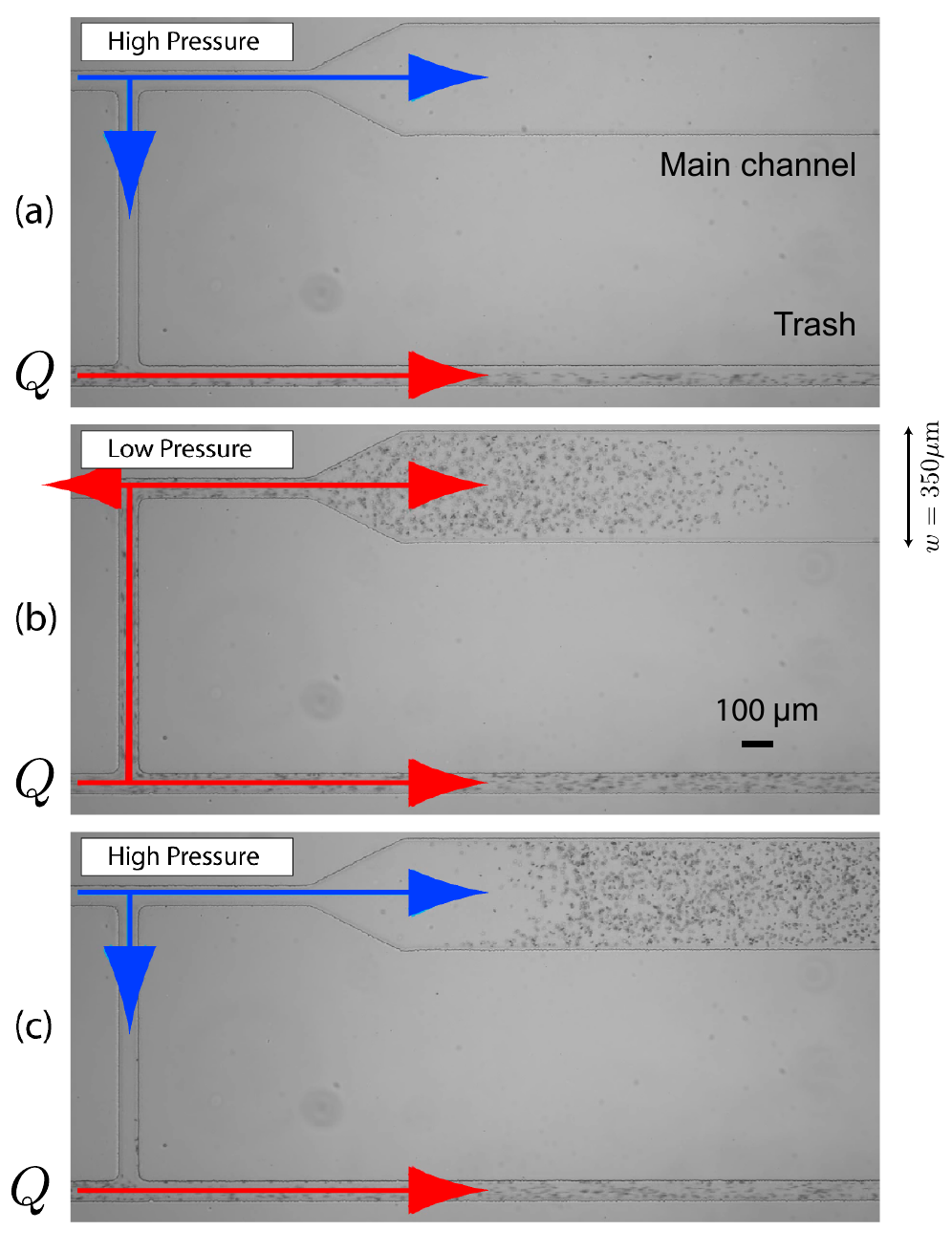}}
\caption{Working principle of the microfluidic commuter. At the bottom inlet, red blood cell suspension is injected at a constant flow rate $Q$, while the pressure at the top inlet connected to a reservoir of suspending medium is varied periodically. In (a), the pressure is high enough so that the suspending fluid enters into the main channel. Lowering the pressure as in (b) let the red blood cells enter into the commuter, and in particular in the main channel, this creating the forefront of the pulse. In (c) increasing back the inlet pressure allows to end this pulse.  \label{fig:commuter}}
\end{figure}

\section{Experimental set-up: the microfluidic commuter}

The experiment consists in generating pulses of a red blood cell suspension at the entrance of a channel and studying their evolution along the channel. In practice, this set-up is based on a standard microfluidics chip whose function and control are detailed below, made of a slab of PDMS  moulded on a SU8 template using standard soft lithography techniques and bonded to a glass slide after plasma treatment. 

The initial dispersion of a pulse is rather fast and of order the velocity difference between fast RBCs at the center of the channel and slow ones near walls. However the long term evolution of the dispersion rate is governed by the migration of RBCs across streamlines, from the walls to the center line. As shown in \cite{losserand19}, this migration is  slow compared to axial velocities and the distance required for RBC centering (in the ideal case of dilute suspensions) can be of order 1000 times the channel thickness.
A long channel is therefore needed to study the time evolution of a pulse. As a consequence,
tracking a single pulse along the channel while keeping a good local resolution is an impossible task due to its rapid stretching and convection.
We instead chose  to generate periodic pulses through a stable and repeatable process in order to study their profile at different axial positions along the channel.

Usually, inclusions of a fluid inside another one (e.g. bubbles or drops) are produced by microfluidic flow-focusing devices. The non-miscibility of the two fluids and the existence of a surface tension are the key ingredients that allow the spontaneous generation  of well separated and periodic inclusions.  Here, the separation process between the suspension and the particle-free fluid requires to actively control the alternation of pure suspending fluid and RBC suspension.
In practice, a direct control of pulse production through a flow focusing system or T-junction (as is done for bubbles or drops) is not possible through direct flow rate control, due to the lag times of flow-controlled microfluidics, and our attempts to control the experiment only through imposed pressures at the inlets have revealed that stabilization is not easily achieved.

In order to retain the fast response times of pressure control, 
we opted for a mixed approach, where the total flux of red blood cells  in the microfluidic chip is fixed and imposed by a syringe pump through a secondary channel while part of it is periodically redirected to the entrance of the main channel thanks to pressure pulses imposed by a pressure controller. This solution as the advantage of being simple in its design and manufacturing process, in contrast with more complex valve systems \cite{Unger13,Baek05}.

More precisely, as sketched in Fig. \ref{fig:commuter}, a fixed flux $Q$ of RBC suspension is imposed  at one inlet of 
a H-shaped commuter. The other inlet is connected to a reservoir of suspending fluid whose pressure is controlled. For low enough pressures, the RBC suspension is diverted into the main channel. For larger pressures, it flows through the secondary (parallel) channel leading to a waste outlet, 
while the suspending fluid enters the main channel, thus separating the previous pulse from the next one.

\begin{figure}[t]

\resizebox{8.6cm}{!}{
  \includegraphics{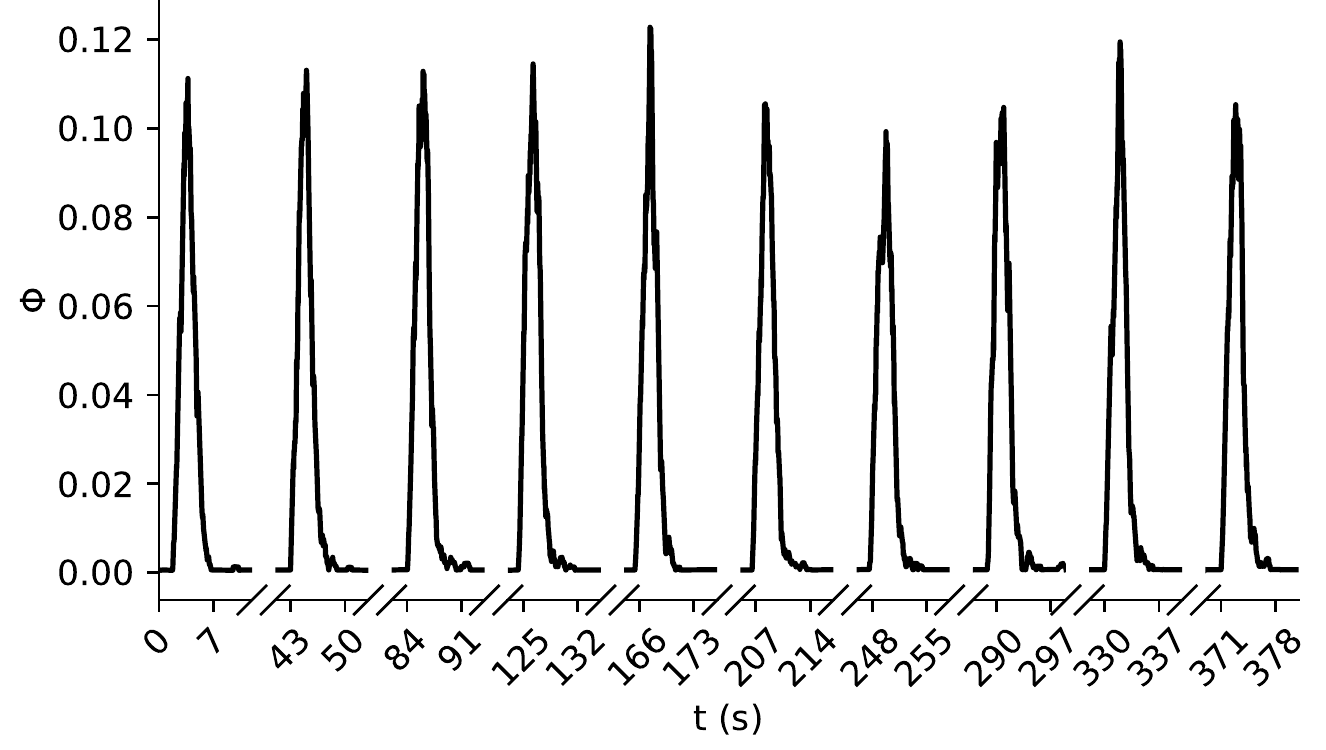}}
\caption{Sequence of temporal profiles of a series of pulses, right after the commuter. \label{fig:sequence}}
\end{figure}

The pressures and flow rates are set such that the initial length of pulses is around 5~mm, with maximal cell velocity of $1.5$ mm.s$^{-1}$. This results in pulses of initial temporal length of a few seconds, separated by intervals of around 30~s, as seen in Fig. \ref{fig:sequence} which illustrates the repeatability of the pulse generation process,
 {with fluctuations of pulse length smaller than 5\%.}

 The main channel is a long serpentine channel of rectangular cross section $2 w \times 2 h$ = 350$\times$ 33 $\mu$m (except for the data of Fig. \ref{fig:conc} where $2h=40\,\mu$m). The $x$ coordinate corresponds to the flow direction while the cross-section is defined by $-w<y<w$ and $-h<z<h$. In practice, as $w \gg h$ there is almost no shear in the $y$ direction and the flow is quasi 2D. We therefore assume there is no significant cell motion nor velocity variation in the $y$ direction. The focal plane of the bright field microscope is $Oxy$.  The RBC velocities and concentrations are evaluated in the $-w/2\le y\le w/2$ area. At each $x$ position along the main channel, 8 to 10 pulses are considered in order to average out the fluctuations of pulses; for each of them successive pictures  are taken at 30 fps thanks to a monochrome camera mounted on a IX71 Olympus inverted microscope with a motorized stage. A blue filter in the illumination beam (434 nm $\pm$17 nm) corresponding approximately to the absorption peak of hemoglobin at 410 nm \cite{Wojdyla12}  is used to enhance the contrast. For each individual picture, taken at time $t$, the mean volume fraction and the mean maximal velocity are computed on the whole field of view (of length 470 $\mu$m in the $x$ direction).

The local volume fraction of RBCs $\Phi$ is determined following the Beer-Lambert law of absorption, which is generally considered to be relevant in micrometric channels up to hematocrits (RBC volume fraction) of around 20 \% \cite{grandchamp13,roman16,Shen16}. The absorption coefficient was determined by a calibration with images at low volume fraction, where a direct measurement can be made by counting individual cells. The reference intensity, assumed to be that with no cells, is taken by considering the minimal intensity among all pixels in the channel, which we checked to be accurate enough in the considered range of volume fractions.

In addition, routines written in Python with the OpenCV library \cite{Opencv_library} determines the centre of mass of each cell. Thanks to an acquisition frequency of 54 fps, the displacement between two frames of RBCs flowing at maximum at 1.5 mm/s can be determined by a tracking routine. We denote by  $u_0$ the maximal velocity of the RBCs.

Blood samples are provided by the \'Etablissement Français du Sang (EFS Rhône-Alpes) from healthy donors. RBCs were separated by centrifugation after being washed three times in a solution of phosphate-buffered  saline  (PBS  tablet from Sigma). To prevent sedimentation of RBCs in channels, the RBCs were re-suspended in a density matching PBS solution made of a 65/35 V/V mixture of water and iodixanol solution (Optiprep from Axis-Shield) \cite{roman12,losserand19}. This suspending fluid  has a density of $1.112 \pm 0.001$ g/mL, which almost prevents the sedimentation of the RBCs and a viscosity $\eta_0$  of  $1.9$ mPa.s at 20$^\circ$C which is a little higher than plasma viscosity (1.54 mPa.s at 25$^\circ$C  \cite{Lawrence50}).

Density-fractionated RBCs were prepared by centrifugation in discontinuous density gradients with Optiprep and PBS mixtures as described in \cite{losserand19}, the 4-layer gradient leading to the separation of a top (light) fraction of RBCs of density $1.103 \pm 0.003$ g/mL, and a bottom (heavy) fraction of density $1.123 \pm 0.005$ g/mL.  As these two fractions represent only respectively 5\% and 10\% of the total hematocrit, this separation confirms that the average density in the sample is $1.112$ g/mL.

\begin{figure}[t]
\centering
\resizebox{\linewidth}{!}{
  \includegraphics{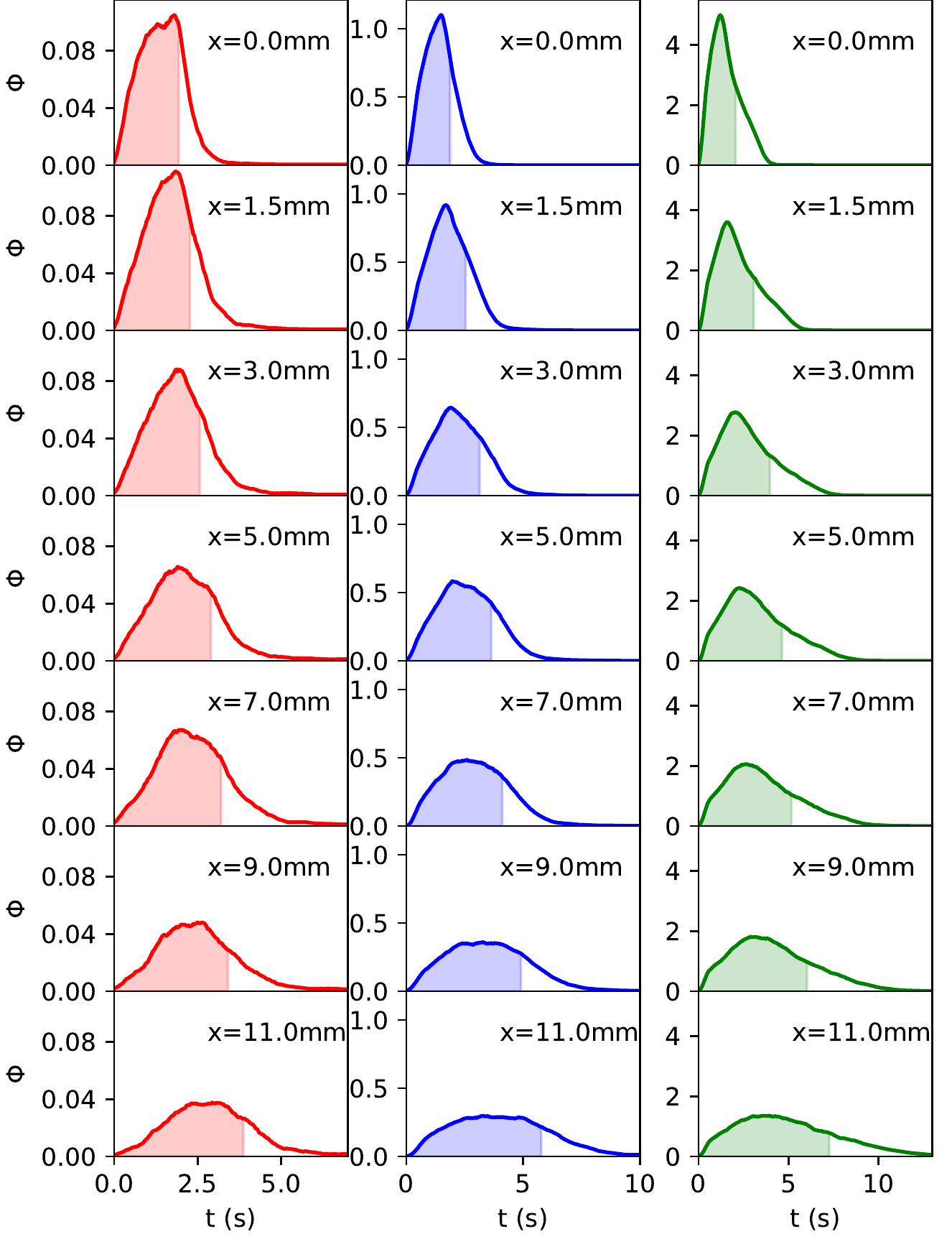}}\caption{Temporal profiles of pulses for different $x$ positions (from top to bottom, 0, 3, 7, 11, 15, 20 et 26 mm) and different initial peak hematocrits (Left: 0.1 \% ; middle 1.1 \% ; right 5\%). Shaded areas correspond to the first 80 \% of cells. \label{fig:profiles}}
\end{figure}

\section{Model}

The evolution of a pulse length $X(x)$ along the channel (or equivalently its duration $\tau (x)$ at a given coordinate $x$) is a consequence of the dispersion of RBC velocities in the non-uniform channel flow profile: as depicted in Fig. \ref{fig:expl} the first cells  of a given pulse reaching position $x$ are those flowing at maximum velocity, i.e. cells that have been at $z=0$ in the middle plane of the channel from the entrance. The tail of the pulse is made of the slowest cells, which are the ones responsible for the increase of $\tau(x)$, that is those initially located close to the walls where velocity is minimum and then slowly migrating towards the center line until all cells eventually reach the same terminal velocity and $\tau(x)$ ceases to increase.  In the absence of transverse migration of these cells, the increase of pulse time duration $\Delta \tau (x) =\tau(x)-\tau(0)$ would be a linear function of $x$ of slope $1/\Delta V$, where $\Delta V$ is the velocity difference between the fast and the slow cells. If slow cells experience a lateral migration towards the center, the growth of $\tau(x)$ (or $X(x)$)will be sublinear. 

We have shown in  Ref. \cite{losserand19} that in the semi-confined case considered here, the velocity of the RBCs can be well approximated by the unperturbed flow velocity averaged over the cell extension. More precisely, we consider $R_0$ the effective radius of a cell defined by  $R_0=(3 \mathcal{V}/ (4 \pi))^{1/3}$, where $\mathcal{V}$ is the volume of a cell. For $\mathcal{V}=90$ $\mu$m$^3$ \cite{McLaren87}, this leads to $R_0=2.8$ $\mu$m. In our channel geometry the longitudinal cell velocity $u_x(z)$ is then 
\begin{equation}
    u_x(z)=\frac{1}{2R_0}\int_{z-R_0}^{z+R_0} v(Z) dZ,
\end{equation}

where $v(Z)$ is the (unperturbed) fluid velocity at position $Z$. Assuming a parabolic profile for the quasi-2D flow, one obtains 

\begin{equation} \label{eq:ux}
    u_x(z)=u_0 \left(1- \frac{3 z^2}{3 h^2-R_0^2}\right),
\end{equation}

which is a parabolic flow field in a channel of width $2h$ with negative slip length $\sqrt{h^2-R_0^2/3}-h$ and where $u_0=v(0)(1-R_0^2/3h^2)$ is the velocity of the centered cells.  {Note that in our experimental cases the ratio $R_0/h$ is about 0.16 or lower, which leads to corrections on the velocity of 1-3 \% only depending on the $z$ position (Eq. \ref{eq:ux}) compared to assuming that the RBC velocity is the unperturbed flow velocity at the position of the RBC center, and thus a similar correction on subsequent results. Nevertheless, we chose to keep this correction in the model for more generality.}

Let us consider a slow cell of coordinates $(x_s,z_s)$, entering the channel at $(0,z_0)$ with $-h<z_0<0$. Its trajectory is determined by the longitudinal velocity $u_x(z_s)$ (Eq. \ref{eq:ux}) and its transverse velocity $u_z(z_s)$. In Ref. \cite{losserand19}, we showed that the transverse migration of a cell can be described by the following scaling law: 
\begin{equation} \label{eq:uz}
   u_z=\xi \frac{R_0^{\delta+1} \dot{\gamma}(z_s)}{(z_s +h)^\delta}.
\end{equation}
where $\xi$ is a dimensionless migration amplitude, $\delta$ a migration exponent (usually between 1 and 2) and $\dot{\gamma}=dv/dz=-2 v(0) z/h^2=-6 u_0 z/(3h^2 -R_0^2)$ is the local shear rate in the particle free fluid. 

Equations \ref{eq:ux} and \ref{eq:uz}, combined with initial conditions, contain in principle all the ingredients required to predict the evolution of the pulse length and shape, at least in the dilute limit where transverse migration is the sole mechanism of RBC motion across streamlines. As there is no general analytical solution for $z_s(x_z)$ and $X(x)$ to these equations, we first derive the asymptotic limit of pulse length as $x \to \infty$ and the behavior of $X(x)$ in this limit.

\subsection{Asymptotic pulse length}

The time derivative of the pulse length $X$ is $dX/dt=u_0-u_x(z_s)$ where $u_x(z_s)$ is the velocity of the slowest cell. We can then write:
\begin{equation}
\label{eq:dXdzs}
    \frac{dX}{dz_s}= \frac{dX}{dt} \frac{dt}{dz_s}=\frac{u_0-u_x(z_s)}{u_z(z_s)}=-\frac{z_s(z_s+h)^\delta}{2 \xi R_0^{\delta+1}}
\end{equation}

This yields the length of the pulse when the slowest cell has migrated from $z_0$ to $z_s$: 

\begin{equation}
    X({ z }_{ s })=X_0+ A({z}_{0})+\frac { h^{ \delta +2 }\left( 1+\frac { z_s }{ h }  \right) ^{ \delta +1 }\left( 1-(\delta +1)\frac { z_s }{ h }  \right)  }{ 2\xi R_{ 0 }^{ \delta +1 }(\delta +1)(\delta +2) } 
\end{equation}

where $X_0$ is the initial pulse length and $A(z_0)$ is a constant that depends on the initial position $z_0$ of cells that are closest to the walls:
\begin{equation}
    A(z_0)=-\frac { h^{ \delta +2 }\left( 1+\frac { z_0 }{ h }  \right) ^{ \delta +1 }\left( 1-(\delta +1)\frac { z_0 }{ h }  \right)  }{ 2\xi R_{ 0 }^{ \delta +1 }(\delta +1)(\delta +2) }
\end{equation}

Note that $A(z_0)=0$ in the ideal case where $z_0=-h$ (which is not possible due to the finite size of cells).
The asymptotic pulse length is obtained when the slowest cells reach the center ($z_s=0$) and the corresponding pulse duration is simply $\tau_\infty=X_\infty/u_0$:
\begin{equation}\label{eq:Xinf}
    X_\infty=X_0+A(z_0)+\frac{h^{\delta+2}}{2 \xi R_0^{\delta+1} (\delta+1)(\delta+2)}
\end{equation}

Interestingly, equation \ref{eq:Xinf} shows that in a sufficiently long channel and by a proper control of initial conditions $(z_0,X_0)$, it is in principle possible to relate the final pulse length $X_\infty$ to  migration parameters $\xi$ and $\delta$ through a simple scaling law. Simply measuring the macroscopic parameter $X_\infty$ in channels of different thicknesses $h$ is therefore sufficient to derive microscopic parameters $\xi$ and $\delta$ that are intrinsic characteristics of the cell mechanical properties in confined flow.

\subsection{Asymptotic behavior}

For a channel with a finite length, for which the asymptotic pulse length cannot be sufficiently approached before the exit, it is necessary to analyze the behavior of $X(x)$ along the channel. As no exact analytical solution for $z_s(x)$ and $X(x)$ can be derived from equations \ref{eq:ux}, \ref{eq:uz} and \ref{eq:dXdzs}, one can look for the asymptotic behavior as $z_s \to 0$ i.e. for $\vert{z/h}\vert \ll 1$. In this limit, expanding the previous equations at order 1 in $z$ one gets:
\begin{eqnarray}
\frac{dX}{dz_s}\simeq-\frac{1}{2\xi(R_0/h)^{\delta+1}}\frac{z_s}{h} \label{eq:dXdz-1} \\
\frac { dz_s }{ dx }=\frac{u_z}{u_x}\simeq-\frac{6\xi R_0^{\delta+1}}{h^\delta(3h^2-R_0^2)}z_s \label{eq:dzdx-1}
\end{eqnarray}

Equation \ref{eq:dzdx-1} yields the following form for $z_s(x)$:
\begin{equation}
    z_s(x)=z_1 \exp{ \left(-\frac{6\xi R_0^{\delta+1}}{h^\delta (3h^2-R_0^2)}(x-x_1)\right)} 
\end{equation}
where $z_1$ is a transverse position of the slow cells in the channel at which the first-order approximations of Eqs. \ref{eq:dXdz-1} and \ref{eq:dzdx-1} start to be acceptable and $x_1$ the corresponding axial coordinate.
When the tail of the pulse reaches position $x$, the pulse length $X(x)$ is then approximately described by:

\begin{eqnarray}\label{eq:dXdz}
\frac{dX}{dx}&=&\frac{dX}{dz_s}\frac{dz_s}{dx}\nonumber \\&=&\frac{3z_1^2}{(3h^2-R_0^2)}\exp{ \left(-\frac{12\xi R_0^{\delta+1}(x-x_1)}{h^\delta (3h^2-R_0^2)}\right)} 
  \end{eqnarray}
or

\begin{equation}\label{eq:Xasympt}
    X(x)-X_\infty=-\frac{z_1^2 h^\delta}{4  \xi R_0^{\delta+1}} \exp\left(-\frac{12 \xi R_0^{\delta+1}(x-x_1)}{h^\delta (3h^2-R_0^2)}\right)
\end{equation}

Equations \ref{eq:dXdz} and \ref{eq:Xasympt} show that the pulse length should converge to its asymptotic value following an exponential law with a characteristic length scale $L$ that does not depend on the initial positions $z_0$ or $z_1$ but only on RBC parameters and channel thickness:
\begin{equation}
    L=\frac{h^\delta (3h^2-R_0^2)}{12 \xi R_0^{\delta+1}}
    \label{eq:L}
\end{equation}

Equation \ref{eq:L} yields a simple relation between the macroscopic dispersion phenomenon and microscopic parameters $\delta$ and $\xi$, the latter one being directly dependent on cell mechanical properties as seen in our previous study on migration \cite{losserand19}. 

Note that in the dilute limit, which is always reached in a sufficiently long channel where a directed force drives particles towards the centerline, the behavior of a pulse is not diffusive. This contrasts with the Taylor-Aris dispersion where the underlying mechanism of lateral displacements is Brownian diffusion and yields a $x^{1/2}$ scaling for the pulse length.

In cases where the initial condition $z_0$ is such that the condition $\vert{z/h}\vert \ll 1$ is not fully satisfied, it may be useful to extend the asymptotic expansion to the next order. In that case (keeping order 2 in $z_s/h$) one gets:

\begin{eqnarray}
\frac{dX}{dz_s}\simeq-\frac{1}{2\xi(R_0/h)^{\delta+1}}\frac{z_s}{h}\left(1+\delta\frac{z_s}{h}\right) \label{eq:dXdz2} \\
\frac { dz_s }{ dx }=\frac{u_z}{u_x}\simeq-\frac{6\xi R_0^{\delta+1}}{h^\delta(3h^2-R_0^2)}z_s \left(1- \delta \frac{z_s}{h}\right) \label{eq:dzdx}
\end{eqnarray}

Assuming that the approximation is valid from the entrance of the channel where $z_s(0)=z_0$, this yields:

\begin{equation}
    z_s(x) \simeq \frac{h z_0}{\delta z_0 + (h-z_0) e^{x/2L}}\label{eq:zs2}
\end{equation}

And the pulse length is:

\begin{equation}
\label{eq:Xasympt2}
    \begin{split}
    X(x) \simeq & X_0 + \frac{6 h^2 L }{(3 h^2-R_0^2) \delta^2} \Biggl[\frac{\delta z_0}{h}    \frac{\left( 1- e^{-x/2L}\right) \left(\frac{\delta z_0}{h}-1\right)}   {1-\frac{\delta z_0}{h}(1 - e^{-x/2L})} \\
    &- \ln \left(1-\frac{\delta z_0}{h}(1 - e^{-x/2L})\right)\Biggr]
    \end{split}
\end{equation}


\subsection{Full form}

More generally, for a given set of parameters $(z_0,\xi,\delta)$, Eqs. \ref{eq:ux} and \ref{eq:uz} can be solved numerically to recover the exact solution, and for each position $x$, the time $t_s(x)$ needed for the cell to reach position $x$, defined as $x_s(t_s)=x$, can be determined. From it, the time lag $\Delta \tau$ which is the difference between the time needed by the slowest and fastest cells to reach $x$, respectively, can be derived:
\begin{equation}
    \Delta \tau (x) = t_s(x)  - \frac{x}{u_0}. \label{eq:timelag}
\end{equation}

Note that for the sake of comparison with experiments, $\Delta \tau$ is indeed the difference between the pulse durations at position $x$ and at the entrance: $\Delta \tau=\tau(x)-\tau(0)$. These quantities are directly measured in the experiment, and can be converted to spatial length via $X(x)=u_0 \tau(x)$ and the corresponding pulse elongation is:
\begin{equation}
    \Delta X (x) = u_0  \Delta \tau (x) = u_0 t_s(x) - x. \label{eq:FF}
\end{equation}

To analyse experimental data using this model, an optimization procedure as a function of the unknown parameters $(z_0,\xi,\delta)$ was implemented. We showed in Ref. \cite{losserand19} that within such a procedure the two parameters $\xi$ and $\delta$ are strongly correlated and a continuum of $(\xi, \delta)$ pairs can yield good agreement with a measured migration trajectory. However, an analysis run with channels of varying height showed that a satisfactory universal migration law is obtained by choosing $\delta=1.3$. We stick to this choice in this study.

The optimization follows a differential evolution method \cite{Storn97} which is implemented using \textit{lmfit} library in Python \cite{Newville14}. The quality of the fit and the parameter values obtained by this optimization method and by fitting with the asymptotic forms of equations \ref{eq:Xasympt} and \ref{eq:Xasympt2} are compared in the following, these asymptotic forms having the advantage of allowing a simpler, direct fitting procedure.

\section{Results and discussion} \label{sec:results}

\subsection{General behavior}

The experimental data consists, for each position $x$ considered along the channel, in a measurement of the temporal profile $\Phi(t,x)$ of the pulse flowing through position $x$. We take as a volume fraction reference the maximum value of the hematocrit $\Phi_p$ within the pulse. 

As shown in Fig. \ref{fig:profiles}, temporal profiles are stretched as the pulse travels along the channel and we characterize the evolution of a pulse by measuring its temporal length $\tau(x)$. In practice, it is taken as the time after which $80\%$ of the cells have passed position $x$. This choice provides more robust results, that are insensitive to outliers in the tail of pulses (see Supplemental Material).
To remove the influence of the initial pulse length, we consider the increase in temporal length $\Delta \tau (x) = \tau(x)-\tau(0)$, where $x=0$ corresponds the entrance of the main channel.

\subsection{Effect of cell deformability}

To establish a link between dispersion and individual cell dynamics, we consider in this section a dilute limit with an initial peak concentration of around  0.1\%, for which cell-cell interactions can be neglected.

Red blood cells have a timelife of around 4 months. As they get older, cells become denser and less deformable \cite{linderkamp82}, which eventually leads  to their elimination in the spleen \cite{Deplaine11}. Sorting the cells by density amounts then to sort them by deformability, a generic notion that is indeed  physically related to several properties such as cytosol viscosity, membrane shear or bending elasticity \cite{Bocci81,Waugh92}.

\begin{figure}
\resizebox{\linewidth}{!}{\includegraphics{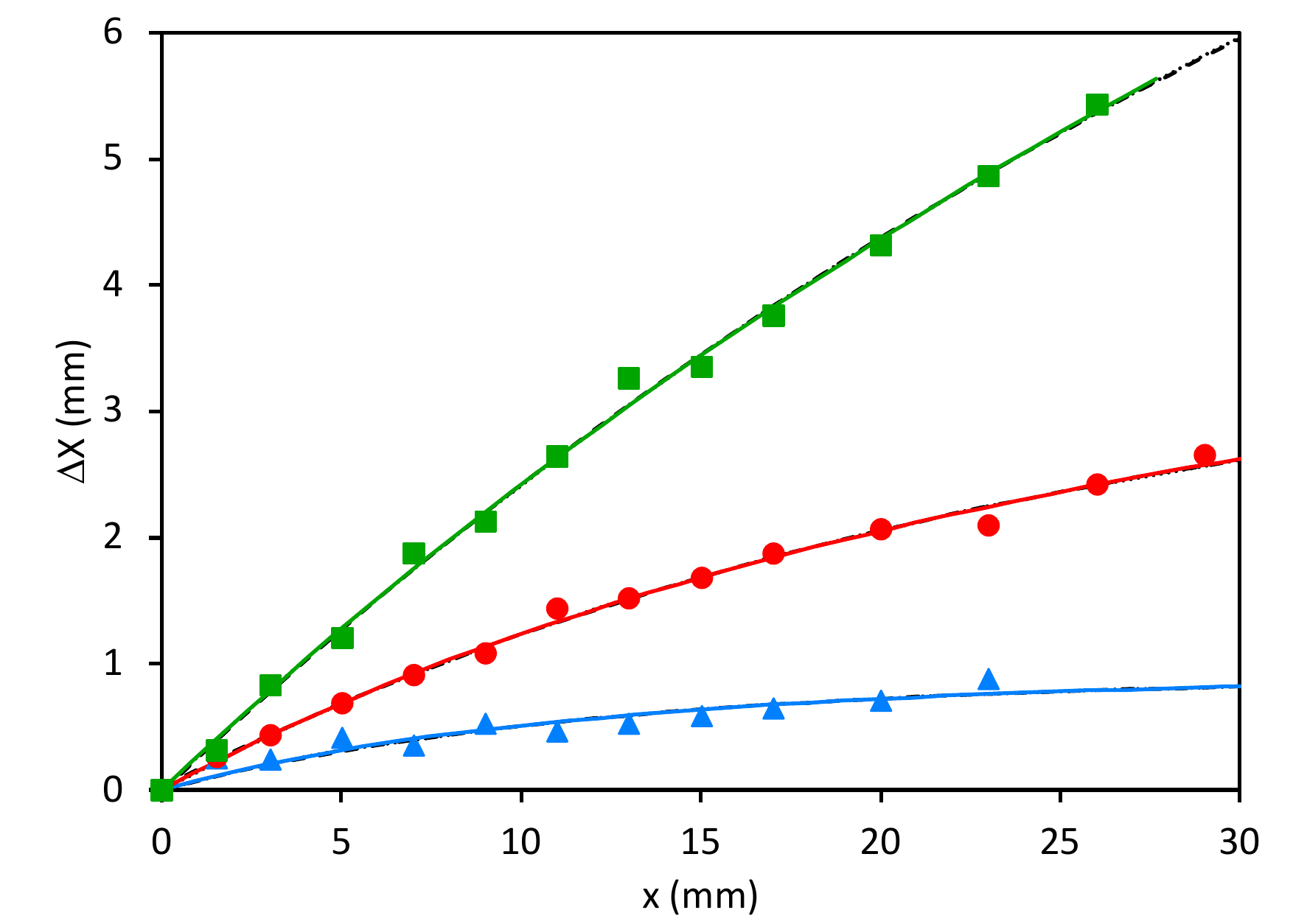}}
  \caption{Evolution of pulse length for cells with different deformabilities, as assumed from their sorting by mean density $d$ ($\blacktriangle$: $d=1.103$ ; $\bullet$: $d=1.112$ ; $\blacksquare$: $d=1.123$), in a channel of thickness $2h=33$~$\mu$m. Lines show fits of the models with $\delta=1.3$ (colored line: Eq. \ref{eq:FF} (FF), dashed line: Eq. \ref{eq:Xasympt} (AF), dotted line: Eq. \ref{eq:Xasympt2} (AF2). Note that FF, AF and AF2 are almost superimposed.). \label{fig:deform}}
\end{figure}

Fig. \ref{fig:deform} shows the evolution of  
pulse length $\Delta X(x)=X(x)-X_0$
as a function of travelled distance $x$, for three mean densities $d$ of RBCs. As in Ref. \cite{losserand19},  mean density  $d=1.103$ g/mL corresponds to the lightest cells, having densities in the range 1.099-1.106 g/mL, while mean density $d=1.123$ g/mL corresponds to the heaviest cells, with densities in the range 1.118-1.128 g/mL. Results labeled as $d=1.112$ g/mL correspond to the whole, unsorted population of cells in the blood sample. 

As illustrated in Fig. \ref{fig:profiles}, 
the pulse length
grows as it travels along the channel. A faster growth indicates a slower migration of the slowest cells, but the behavior of the $\Delta X(x)$ also depends on the initial condition $z_0$. A fit of the experimental curves with the full form (FF) of Eq. \ref{eq:FF} and the asymptotic forms  of Eqs. \ref{eq:Xasympt} (AF) and \ref{eq:Xasympt2} (AF2) with fixed $\delta=1.3$ yields the  couples of parameters  $(z_0, \xi)$ for FF and AF2 and $(L, \xi)$ for AF summarized in Table \ref{tab:density}.
 {Note that a direct measurement of $z_0$ is not possible in the experiment: here the optical axis is the $z$-axis and a direct observation along the $y$-axis (the vertical direction in Fig. \ref{fig:commuter}) is impossible due to the small  thickness of the main channel compared to its $x$ and $y$ dimensions. However the obtained $z_0$ values are physically consistent as explained in the following. Also note that the AF model does not need $z_0$ as explicit fitting parameter.}

The fitted values of $z_0$ show that in all cases cells are already quite far from  channel walls  (here $h=16.5$~$\mu$m) at the entrance of the main channel, as a result of their initial displacement in the inlet channels where the pulse is created. Unsurprisingly, the more deformable (less dense) ones that migrate faster are already the closest to the centerline at $x=0$. This is indeed confirmed by the fact that the exponential asymptotic law (AF) and the 2nd order asymptotic law (AF2) fit very well with the experimental data and are almost superimposed with the full form (FF) modeling. 

 {The relaxation distance $L$ is in the range of several centimeters  here, which is about 1000 times the channel thickness. This is obviously much longer than the convergence length of the cell-free layer (CFL) observed in other studies on more concentrated suspensions \cite{katanov15} where a balance between migration and shear-induced diffusion takes place. This comparison was discussed in our previous study \cite{losserand19}.}

\begin{table*}[t]
\centering
\caption{\label{tab:density} Fitted values of parameters when varying RBC mean density, for the full form optimization (FF) and the 1st order (AF) and 2nd order (AF2) asymptotic forms. For AF and AF2, $\xi$ values (in italics) are derived from the fitting parameter $L$. For the sake of comparison, $X_\infty$ was also extrapolated from the FF and AF2 fits (italics). The results of the direct measurements of Ref. \cite{losserand19} are recalled.}
\begin{tabular}{c|c c c|c c c c|c c c c|c}
  \cline{2-13}
{} &  \multicolumn{3}{c}{FF} & \multicolumn{4}{c}{AF} & \multicolumn{4}{c}{AF2} &  Ref. \cite{losserand19}\\
\hline
$d$   & $z_0$  & $\xi$ & $X_\infty$  & $X_\infty$ & A & $L$   & $\xi$ & $z_0$ & $L$   & $\xi$ & $X_\infty$ & $\xi$\\
(g/ml) &  ($\mu$m)  &  ($ 10^{-3}$) &  (mm)   &  (mm)  &  (mm) &  (mm)  &  ($ 10^{-3}$) &  ($\mu$m)  &  (mm) &  ($ 10^{-3}$) &  (mm) & ($ 10^{-3}$)\\
\hline
1.103   &  -4.5 & 15.3 & \textit{0.92} & 0.89 & 0.89  & 11.7  & \textit{20.7}  & -4.6 & 14.3 & \textit{16.9} & \textit{0.90} & 14\\
1.112   &  -6 & 5.2 & \textit{4.37} & 3.81 & 3.79  & 25.9  & \textit{9.3}  & -6.4 & 34.5 & \textit{7.0} & \textit{3.95} & 11\\
1.123   &  -7.6  &  1.7 & \textit{18.7}  & 12.4 & 12.4 & 45.8  & \textit{5.3}  & -8.5 & 75.1 & \textit{3.2} & \textit{14.1} & 6.5\\
\hline
\end{tabular}
\end{table*}

The tendency observed for the migration amplitude $\xi$ and the relaxation distance $L$ show that lighter cells migrate faster and the corresponding pulse length stabilizes after a shorter distance (about 20 mm vs nearly 50 mm for heavier cells). 
These differences are a strong marker of dispersion in mechanical properties within a given sample, as also described for cells under simple shear flow \cite{fischer13,minetti19}. Note that, the values of $\xi$ fitted with the full-form model (FF) exhibit much more marked differences when varying RBC density than the asymptotic forms: The optimization procedure of the FF model may be less selective on the $(z_0,\xi)$ couple (i.e. a range of couples may give equivalently satisfactorily results) while the experiment does not allow to check if the obtained $z_0$ values are correct. 
On the other hand, in the asymptotic forms, the characteristic length $L$ is a more explicit 
feature of the experimental curve and may therefore be more robust way of estimating $\xi$ in addition to being computationally easier to implement, as long as the asymptotic regime is reached. 

Remarkably, the values yielded by all models for $d=1.103$ g/mL are actually very close to those obtained previously by directly measuring migration velocities \cite{losserand19}, even though the blood sample was different. For higher cell densities, two features emerge: (i) there are slightly more marked differences between the FF model and asymptotic models, which is likely due to higher $\vert{z_0}\vert$ values, with the AF2 model obviously providing a better approximation, and (ii) $\xi$ values are lower than in direct measurements \cite{losserand19}. 
This reflects the fact that the dynamics of the back of the pulse is governed by the slowest cells which, possibly after a short transient, are the ones having the lowest transverse migration velocities. As the $d=1.123$ g/mL sample covers a relatively large range of densities ($1.118-1.128$ g/mL), some of the cells are significantly slower that the average. 
Concerning the whole sample $d=1.112$ g/mL, in the direct measurement of \cite{losserand19} the average migration amplitude of the whole population has an intermediate value between light and heavy cells. Here, we would expect the pulse dynamics to be  governed by the slowest cells, which happen to be the same as in the $d=1.123$ g/mL sub-population, and yield the same $\xi$ value. While values are closer than what they are in the direct measurements, there is still a significant difference. However, the way the pulse length is experimentally defined (by considering the 80\% of RBCs at the front of the pulse) removes some of the slowest (and densest) cells from the whole population and explains why the pulse length does not behave exactly as for the denser case $d=1.123$ g/mL.

The deformation of cells under flow can also be monitored externally through the viscosity $\eta_0$ of the suspending fluid. The dynamics of RBCs draws in reality a complex diagram, even in unbounded shear flow, which depends on the shear stress $\eta_0 \dot{\gamma}$, where $\dot{\gamma}$ is the shear rate, and on the viscosity contrast $\lambda$ between the inner and the outer fluid \cite{goldsmith72,bitbol86,dupire12,fischer13,lanotte16,minetti19}. A very rough view is that increasing the viscosity of the external fluid increases both the stress and the relative weight of dissipation mechanisms outside and inside the cell, allowing a transition from solid-like, flipping motion to droplet-like motion where the cell adopts a fixed shape relatively to the flow; the rotational component of the imposed stress is then accommodated by inner fluid rotation instead of a rotation of the whole cell. Because of this, an increase of the migration velocity with the viscosity of the external fluid is expected, as demonstrated in Ref. \cite{grandchamp13} for RBCs in simple shear flow near a wall.

Here, we considered suspending fluids of different viscosities by adding dextran 100 kDa (Sigma) in the suspension while varying the Optiprep volume fraction to adjust buoyancy. Instead of a volume fraction of 35\% in Optiprep, a volume fraction of 33 \% (resp. 31\%) with additional dextran in proportion 32.5 g/L (resp. 50 g/L) lead to a viscosity at 20$^\circ$C of 5.8 (resp. 8.1) mPa.s \cite{audemar2022}. 
 {For the highest viscosity and given that the shear rate in the experiment is below 180 s$^{-1}$, the maximum shear stress is about 1.5 Pa. This is therefore still below the transition to tank-treading \cite{fischer13,minetti19} as was the case in our direct study on migration \cite{losserand19}.}

\begin{figure}
 \resizebox{\linewidth}{!}{\includegraphics{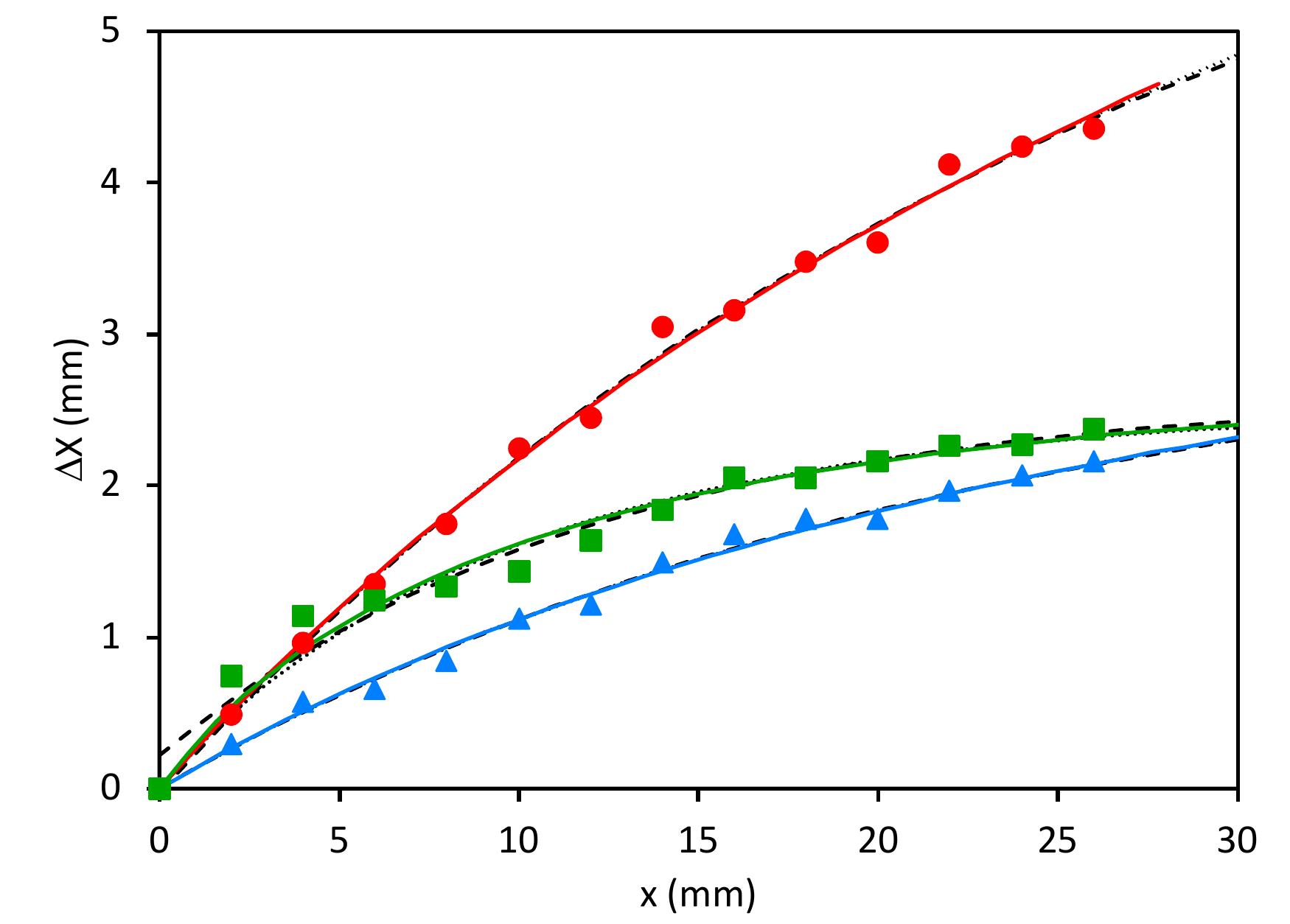}}
  \caption{Evolution of pulse length for cells in fluids of different viscosities $\eta_0=1.9$ mPa.s ($\bullet$), $\eta_0=5.8$ mPa.s ($\blacktriangle$) and $\eta_0=8.1$ mPa.s  ($\blacksquare$), in a channel of thickness $2h=40$~$\mu$m.  Lines show fits of the models with $\delta=1.3$ (colored line: Eq. \ref{eq:FF} (FF), dashed line: Eq. \ref{eq:Xasympt} (AF), dotted line: Eq. \ref{eq:Xasympt2} (AF2))\label{fig:lambda}. }
\end{figure}

The evolution of pulse lengths are shown in Fig. \ref{fig:lambda} and parameter values fitted by FF, AF, and AF2 models are gathered in Table \ref{tab:viscosity}.
The FF fitted value of $\xi$ for $\eta_0=1.9$ mPa.s is close to that obtained in the same experimental conditions in Table \ref{tab:density} for the whole sample ($d=1.112$ g/ml), the small difference being attributable to differences between different blood samples. 
Interestingly, a quick look at the two curves for $\eta_0=5.8$ and $8.1$ mPa.s in Fig. \ref{fig:lambda} does not allow to directly assess the migration velocity of cells (and therefore their deformability) by simply comparing pulse lengths. Only a model-based analysis taking into account the details of the rate of change of pulse length allows to separate the intrinsic property ($\xi$) from the influence of initial conditions ($z_0$). These are in the present case responsible for the $\eta_0=8.1$ mPa.s curve being initially above the other one due to cells closer to the walls at the entrance of the channel and leading to initially faster dispersion, while their asymptotic behavior is in-line with the expected relative values of the lift velocity: the slope of the $\eta_0=8.1$ mPa.s curve becomes smaller than that of the $\eta_0=5.8$ mPa.s curve as cells migrate much faster towards the centerline.

Note that in this series, the ratio $|z_0 |/ h$ is significantly bigger than in the series of Table \ref{tab:density}, which leads to poorer performance of the AF model and bigger differences between the 1st order (AF) and 2nd order (AF2) asymptotic forms.

\begin{table}
\centering
\caption{\label{tab:viscosity}
Fitted values of parameters when varying the viscosity of the suspending fluid, for the full form optimization (FF) and the 1st order (AF) and 2nd order (AF2) asymptotic forms (only the main parameters of interest are shown).}
\begin{tabular}{c|c c|c c|c c}
  \cline{2-7}
{} &  \multicolumn{2}{c}{FF} & \multicolumn{2}{c}{AF} & \multicolumn{2}{c}{AF2} \\
\hline
$\eta_0$   & $z_0$  & $\xi$   & $L$   & $\xi$  & $z_0$  & $\xi$\\
(mPa.s) &  ($\mu$m)  &  ($ 10^{-3}$)   &  (mm)   &  ($ 10^{-3}$) &  ($\mu$m)  &  ($ 10^{-3}$)\\
\hline
1.9   & -9.1 & 5.9   & 28.0  & 16.3 & -10.2 & 10.1 \\
5.8   &  -7.0 & 11.0   & 22.8  & 20.1 & -7.4 & 14.5 \\
8.1   &  -9.9  &  25.4   & 12.0  & 38.2 & -10.5 & 34.6  \\
\hline
\end{tabular}
\end{table}

\subsection{Effect of cell-cell interactions}

\begin{figure}

\resizebox{\linewidth}{!}{\includegraphics{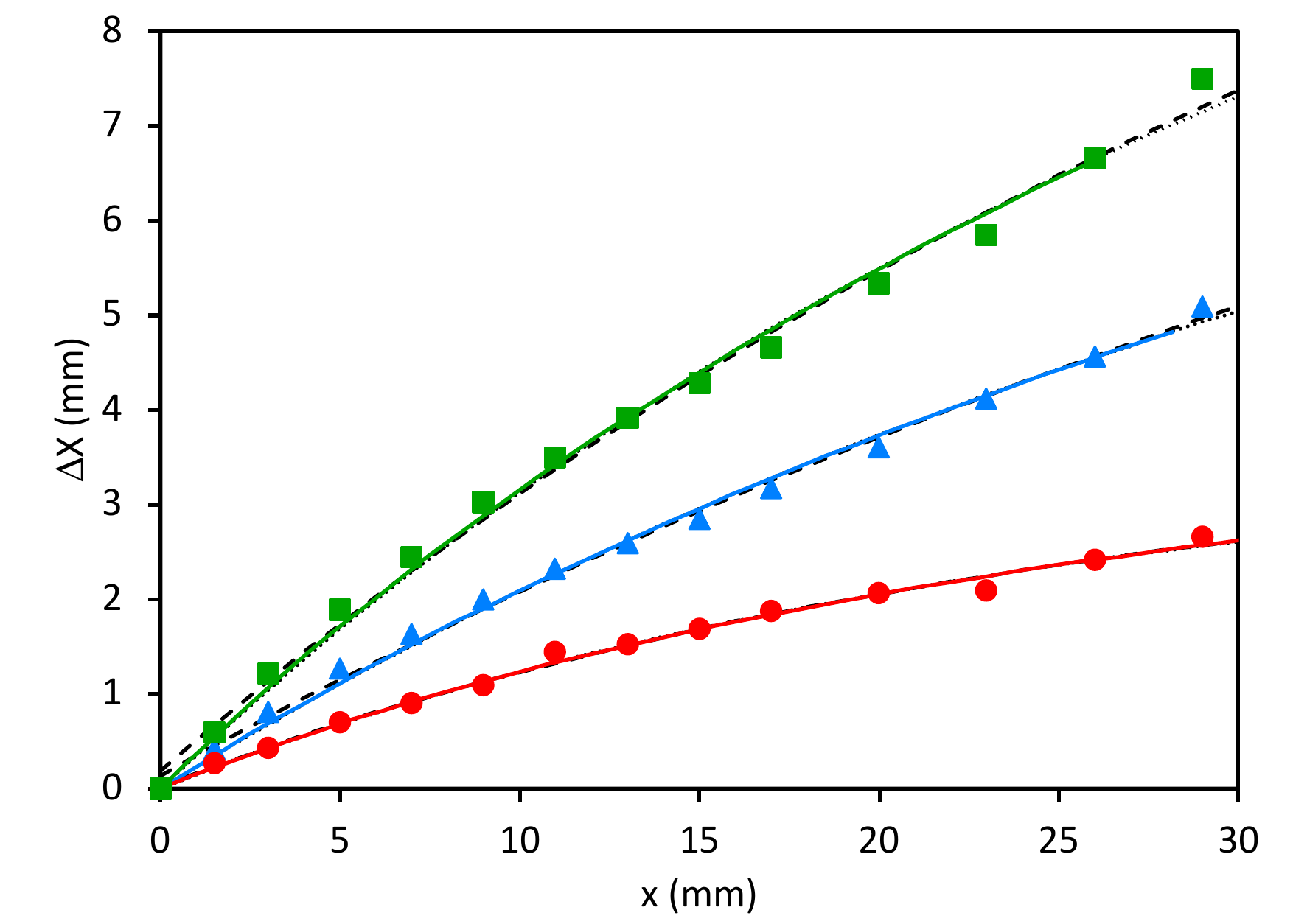}}
  \caption{Evolution of pulse length for cells with different initial peak concentrations (corresponding to the three columns of Fig. \ref{fig:profiles}); $\bullet$: $\Phi_p=0.1\%$, $\blacktriangle$: $\Phi_p=1\%$, $\blacksquare$: $\Phi_p=5\%$. Lines show fits of the models with $\delta=1.3$ (colored line: Eq. \ref{eq:FF} (FF), dashed line: Eq. \ref{eq:Xasympt} (AF), dotted line: Eq. \ref{eq:Xasympt2} (AF2)). \label{fig:conc}}
\end{figure}

Hydrodynamic collisions between cells lead to transverse shear-induced diffusion that could be expected to slow-down the migration of cells towards the center. Fig. \ref{fig:conc} shows the evolution of the dispersion for pulses of varying initial peak concentrations, $\Phi_p=$ 0.1 \%, 1 \% and  5\%. Fitting the experimental curves with the FF, AF and AF2 models which assume no interactions between cells can still be formally done in order to extract effective migration parameters, and yields the results of Table \ref{tab:phi}. Note that the dataset for $\Phi_p=0.1$\% is the same as the one labeled $d=1.112$ g/mL in Fig. \ref{fig:deform}. 

\begin{table}
\centering
\caption{\label{tab:phi}
Fitted values of parameters when varying the peak hematocrit of the pulse, for the full form optimization (FF) and the 1st order (AF) and 2nd order (AF2) asymptotic forms (only the main parameters of interest are shown).}
\begin{tabular}{c|c c|c c|c c}
  \cline{2-7}
{} &  \multicolumn{2}{c}{FF} & \multicolumn{2}{c}{AF}  & \multicolumn{2}{c}{AF2} \\
\hline
$\Phi$   & $z_0$  & $\xi$   & $L$   & $\xi$ & $z_0$  & $\xi$ \\
(\%) &  ($\mu$m)  &  ($ 10^{-3}$)   &  (mm)   &  ($ 10^{-3}$) &  ($\mu$m)  &  ($ 10^{-3}$)\\
\hline
0.1   & -6.0 & 5.2   & 25.9  & 9.3 & -6.4 & 7 \\
1   &  -7.2 & 2.2   & 59.0  & 4.1 & -8.0 & 3.7 \\
5   &  -8.6  &  2.0   & 46.4  & 5.2 & -9.9 & 4.1 \\
\hline
\end{tabular}
\end{table}

The dilute suspension has a larger $\xi$ as expected, but a quantitative interpretation of the smaller value of $\xi$ found for large volume fractions is delicate: as the pulse evolves and elongates, local concentration decreases and one should eventually recover the dilute case regime. However, the front of the pulse corresponds to cells that are concentrated around the centerline of the channel where the shear rate is lower, and the concentration peak may therefore survive for some time before actual dilution occurs. This may explain why for $\Phi_p=1$\% and 5\% the evolution of the pulse length is almost linear after $x=10$ mm and does not show any sign of saturation over the whole channel length. This suggests that shear-induced diffusion effects are significant as soon as $\Phi_p \ge 1$\% and that migration parameters cannot reliably be derived from such an experiment. A more elaborate modelling would be needed in that case to include the effects of both transverse migration and shear-induced diffusion.
 {Indeed, the model we proposed here, which consists in deriving the pulse length from the ODEs governing the trajectory of the slowest cell, cannot simply be amended to include diffusive effect. A different modelling approach would be needed in which PDEs for the full concentration field need to be considered to model diffusive and convective (migration) fluxes.}

\subsection{Outlook on dispersion in channel networks}

\begin{figure}[t]

\resizebox{\linewidth}{!}{
  \includegraphics{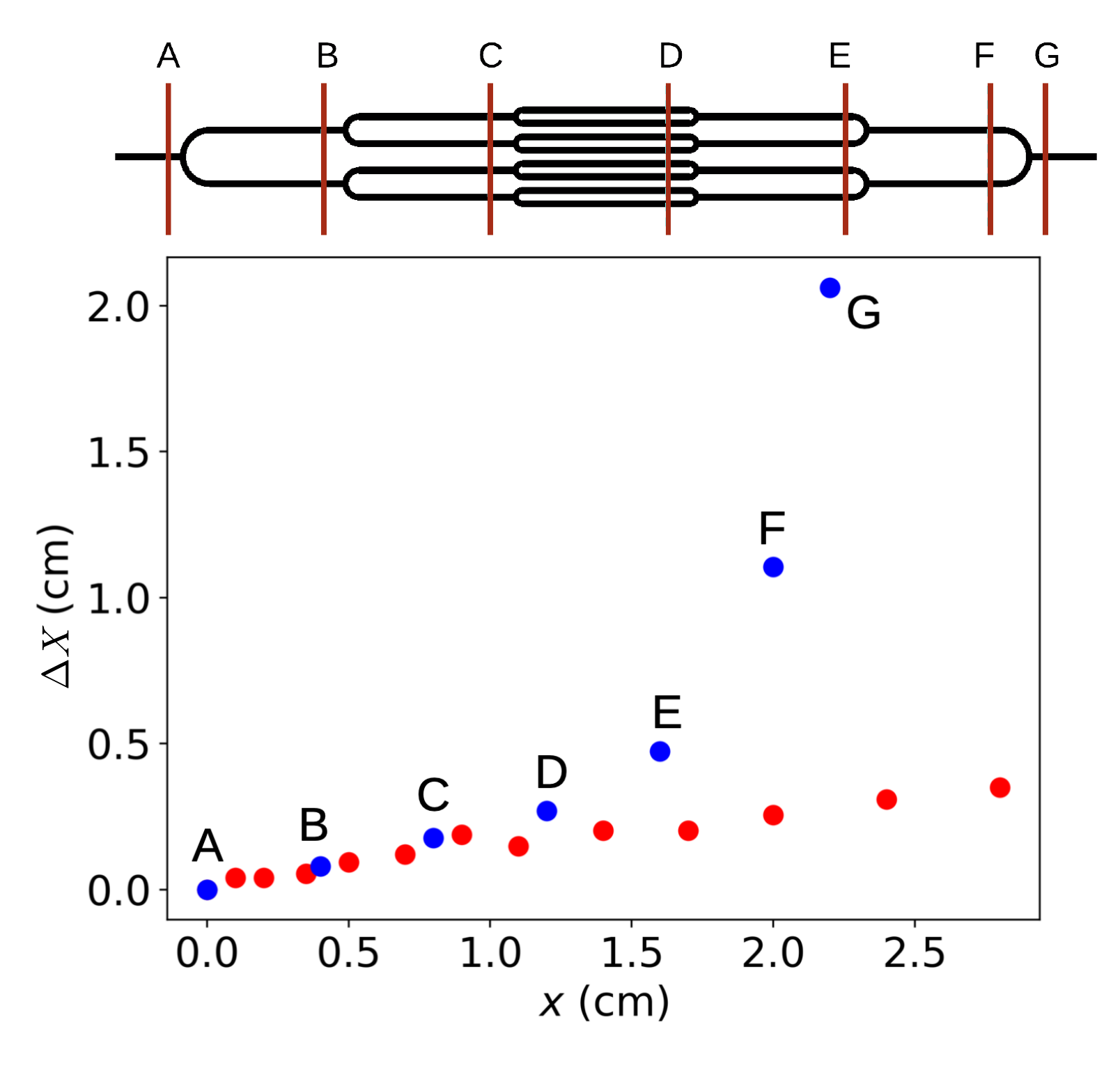}}\caption{Comparison of dispersion between a straight channel and a branched network. The characteristic length  $L$ of a pulse is plotted against longitudinal position $x$ for both configurations. In the branched channel (blue dots), a measuring section was selected right before diverging or converging bifurcation. \label{fig:opening}}
\end{figure}

While the mechanisms discussed in this paper lead to a marked axial dispersion in a single straight channel, the presence of bifurcations in a channel network is responsible for local structural reorganization of the flowing suspension and new initial conditions after each bifurcations that should contribute to this dispersion at the scale of the network.  
As a first step towards a better understanding of this complex question, 
we compared the axial dispersion in two different configurations: a single straight channel and a branched network consisting in a succession of 3 diverging bifurcations followed by 3 converging bifurcations which can be viewed as a very simplified model of microvascular network, as depicted in Fig. \ref{fig:opening}. In both configurations, the individual channel cross section is $2w\times 2h= 30 \times 25\,\mu$m$^2$ and the distance between bifurcations of the network is 4 mm. As the mean flow velocity evolves in the branched network when going through bifurcations, the comparison between the two configurations is based on pulse length $\Delta X=u_0 \Delta \tau $, where $u_0$ is measured at each considered position.

Fig. \ref{fig:opening} shows the evolution of this length as a function of the longitudinal position $x$ along the straight channel or the branched network. It clearly shows that while diverging bifurcations have little effect on the dispersion, converging bifurcations induce a dramatic increase of the dispersion in the branched network, compared to the sole dispersion induced by velocity differences in a straight channel.

\begin{figure}[t]
\resizebox{\linewidth}{!}{
  \includegraphics{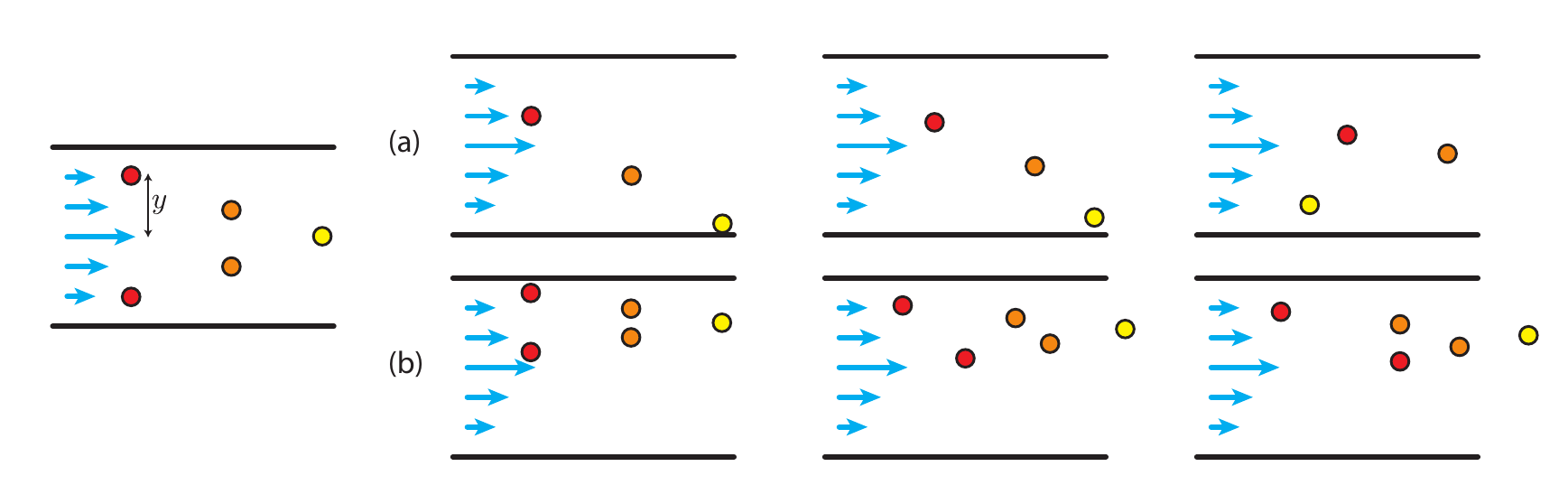}}\caption{Schematic of the behaviour inside the pulse right after (a) a diverging bifurcation and (b) a converging bifurcation. The initial pulse is represented by its front (yellow), intermediate (orange) and rear (red) particles. Right after a diverging bifurcation (a), half of these particles split into the daughter branch ; notably, the front particle is now close to the inner wall and is gradually overtaken by the queue particles. Right after a converging bifurcation (b), particles occupy half the downstream branch and the pulse experiences higher shear rates on average.
  \label{fig:postbif}}
\end{figure}

An interpretation of this result is complicated here by the fact that shear is present in both transverse directions $y$ and $z$ due to the nearly square cross section (resulting in the presence of slow cells near all four lateral walls) while bifurcations mainly redistribute cells in the $y$ direction. Nevertheless, a qualitative interpretation of the asymmetric role played by both types of bifurcation can be proposed through the scheme of Fig. \ref{fig:postbif} where the deformation of an initially very thin pulse is sketched. When going through a diverging bifurcation, front particles that are on the centerline upstream are close to the inner wall downstream of the bifurcation. As a consequences the fastest particles become the slowest ones and this should lead, at least immediately at the entrance of the downstream channels, to a reduction of the dispersion and a contraction of the pulse. Then dispersion as discussed before takes place again, possibly enhanced by the fact that cells have been pushed back to high-shear regions near walls. Depending on channel length between bifurcations, the contraction of the pulse taking place immediately after each diverging bifurcation may significantly compensate the enhanced axial dispersion due to the relocation of particles. In our case, this leads to a dispersion that is very similar or only slightly bigger in the diverging part of the network compared to a straight channel.

After a converging bifurcation, as a first approximation the particle distribution that spans over the whole channel width upstream is squeezed over half the width of the downstream channel. As a result, half of the slow particles are still close to a wall downstream while the other half are located close to the center, with a distance that depends on their initial transverse position, and therefore become fast particles. This  redistribution of the particle cloud in a region where the average shear rate is higher and the average velocity is lower induces a stronger dispersion per unit displacement in the axial direction. This simple picture explains the dramatic increase of the pulse length after each converging bifurcation in Fig. \ref{fig:opening}. 

While this qualitative description of the reorganization of the suspension after both types of bifurcations helps to understand the differences observed in Fig. \ref{fig:opening},  
more data is obviously required to reach a full, quantitative understanding of the involved mechanisms. The distance between each bifurcation, therefore the configuration in which the pulse reaches them, is of course a key parameter here. If RBC migration is fast compared to channel length (narrow channels, very deformable cells with high $\xi$), one can imagine a steady regime in which all cells are centered before the next bifurcation. In such a situation, pulse length should quickly reach an asymptotic value even in a network with bifurcations. In realistic blood micro-circulatory networks however, the high occurrence of bifurcations makes it necessary to take into account the transient distribution of cells \cite{merlo2022}.

\section{Conclusion}

This work reveals that the axial dispersion of Red Blood Cells in channel flow is strongly related to their mechanical properties that control their transverse displacements. In dilute suspensions, our experiments and theoretical modeling show that the axial dispersion dynamics can be directly derived from migration dynamics towards the center of the channel, which is assumed to follow a previously established scaling law. As a consequence a higher cell deformability (or equivalently a higher viscosity of the suspending fluid) decreases the axial dispersion.

Conversely, this general principle and the techniques presented here reveal that with controlled initial conditions, a simple macroscopic measurement of pulse length can be used to easily derive the microscopic migration velocity parameter of a blood sample, which is a direct signature of RBC mechanical properties. Interestingly, this axial dispersion measurement principle, which requires relatively light equipment and much less image and data processing than direct measurements opens up alternative ways to monitor the deformability of RBCs or other cell populations in healthy and pathological situations, e.g. for diagnostic purposes. 

 {Although we observed some variations of the migration parameters obtained in different experiments (pulse dispersion vs. direct migration measurements) or using different fitting models, one could certainly consider an improved experimental setup in which $z_0$ is imposed and the main channel is long enough to allow the pulse to approach its asymptotic length $X_\infty$. This would reduce the number of fitting parameters and decrease the discrepancies between different methods and models as well as allow to use the simpler asymptotic form of the pulse dispersion model.}

Our study of axial dispersion in straight channels represents a first step in the quantitative understanding of hemodynamics in microvascular networks, where the dispersion of RBC transit times, a critical factor in oxygen exchange for instance, is also obviously linked to network topology and bifurcations. Our first study of the axial dispersion in a simple network highlights the starkly different influence of diverging and converging bifurcation and shows a dramatically enhanced dispersion after converging bifurcations. While light is usually shed on the role of diverging bifurcations as they are the locus of uneven distribution in cells within downstream branches, this emphasizes the central role that could be played by the diverging ones regarding the dispersion of the transit time of cells and therefore of the oxygen release within one organ. The contribution of both types of bifurcations remains to be explored through a more systematic study were transit times after cell reorganization would be varied relatively to typical dispersion times.

\section{Acknowledgements}

This work was partly supported CNES (Centre National d'Etudes Spatiales) and the authors thank LabEx Tec 21 (Investissements d'Avenir - grant agreement ANR-11-LABX-0030) for a Proof of Concept fellowship. 



 \bibliographystyle{unsrt}

%

\end{document}